\documentclass{PoS}
\usepackage{amsmath,amsfonts,euscript,amssymb}
\def\be{\begin{equation}}
\def\ee{\end{equation}}
\newcommand{\bea}{\begin{eqnarray}}
\newcommand{\eea}{\end{eqnarray}}

\title{{Condensate mechanism of} \\
{conformal symmetry breaking}}

\ShortTitle{Condensate mechanism of conformal symmetry breaking}

\author{\speaker{Victor Pervushin}%
         \\
        JINR\\
        E-mail: \email{pervush@theor.jinr.ru}}

\author{
    \\ Andrej~Arbuzov, Boris~Barbashov, Alexander~Cherny, Alexander~Dorokhov,
        \\JINR, Dubna,
\\Andrzej Borowiec, \\
ITP, University of Wroc\l aw,  Poland,\\
Rashid Nazmitdinov, \\
JINR,~Dubna~and~Department~de~F{\'\i}sica,~Universitat~de~les~Illes~Balears,~Palma~de~Mallorca,~Spain,
\\Alexander Pavlov, \\
Moscow State Agri-Engineering University, 127550 Moscow, Russia,
\\Vadim Shilin, \\JINR, Dubna
 and Moscow Institute of Physics and Technology, Dolgoprudny, Russia,
\\Alexander Zakharov, \\ITEP, Moscow, Russia.

 }

\abstract{The low energy Gell-Mann--Oakes--Renner relation,
 Higgs particle mass value, and the new observational cosmological data
 are considered as evidence
 of the condensate mechanism of conformal symmetry breaking at the quantum level.
 The condensate mechanism  occurs by means of normal ordering of
 field operators in QCD,  Minimal Standard Model of electroweak
 interactions without the Higgs potential,
 and the Dirac conformal General Relativity with long range forces.}

\FullConference{XXI International Baldin Seminar on High Energy Physics Problems \\
                 September 10-15, 2012\\
                 JINR, Dubna, Russia}

\begin{document}

\section{Introduction}

The symmetry is inseparably connected with conception of beauty. At the same time, intrinsical beauty
is achieved upon condition the symmetry has been breakdown a little and is incomplete.
This incompleteness gives
mysteriousness and attractiveness to the Nature.

In particular, in modern physics  there is a tendency to treat
 the broken  conformal symmetry 
   as the basis of unification of all interactions.
This is the reason to consider
 quantum chromodynamics (QCD), Minimal Standard Model (SM) of electroweak interactions,
and the Dirac version of
General Relativity (GR)
with long range forces \cite{Dirac_73} in the framework of the conformal symmetry principle.
 Our approach to all these theories (QCD, SM, and GR)
 is based on their conformal invariant classical versions and
 the Dirac Hamiltonian scheme in a definite frame of reference \cite{grg}.
 It will be proposed that  conformal symmetry can be broken at the quantum level by the normal ordering of
 field operators, which leads to their condensates and Casimir energies.

\section {Low Energy QCD}

The QCD  description of hadrons is based on the non-Abelian generalization of the
QED  description of bound states \cite{a15,Salpeter}
in concordance with irreducible
representations of the 
Poincar\'e group
\cite{MarkovYukawa,Pervu2}, and
the Dirac gauge invariant quantization \cite{Dirac},
where the potential components are separated  from  radiation ones in a definite
 frame of reference.

 Let us begin from the standard definition of Poincar\'e-invariant (P-inv)
  and  gauge invariant (G-inv) S-matrix elements. Recall that
the S-matrix elements are defined as the evolution operator expectation values
between in- and out- states
 \bea \label{4-2}\underbrace{{{\cal M}_{\rm in,out}}}_{P-inv,G-inv}=
 \underbrace{\langle {\rm out}|}_{P-variant}~
 \underbrace{\hat S[\hat \ell]}_{P-variant,G-inv}~\underbrace{|{\rm in}\rangle }_{P-variant}.
\eea
 The  in- and out- states are the rays defined as the
products of the P-variant   representations  \cite{logunov}
 \be \label{3-2} \langle {\rm out}|=\langle \prod_{J}{{\cal
 P}_J,s_J}~\big|, ~~~|{\rm in}\rangle =\big|\prod_{J }{{\cal P}_J,s_J}\rangle,
 \ee
 where ${\cal P}_J,s_J$ are the total momentum and spin of a state ($J$).
This means that all particles (elementary and composite)  are  far enough
from each other to neglect their interactions in the in- and out- states.
These irreducible representations form a complete set of states. The frames of
reference  are distinguished by the eigenvalues of the appropriate time-like
operator $\hat \ell_\mu=\dfrac{\hat{\cal P}_\mu}{M_J}$ (where $M_J$ is a particle mass).
This operator acts in the complete set  of these states:
 \bea \label{3-2a}
 \hat \ell_\mu|{{\cal P},s}\rangle =\frac{{\cal P}_{J\mu}}{M_{J}}|{{\cal P}_J,s}\rangle .
 \eea
  The Bogoliubov--Logunov--Todorov  rays (\ref{3-2a}) can include bound
states \cite{a15,Salpeter}.

In a definite frame of reference the gauge invariance of the S-matrix elements  (\ref{4-2}) can be  achieved
via the
 Dirac Hamiltonian
approach to QED of 1927.
This approach was based on the constraint-shell action \cite{Dirac}
 \bea \label{6-2} W^{\rm Dirac}_{\rm QED}=W_{\rm QED} \Big|_{\dfrac{\delta
 W_{\rm QED}}{\delta A^{\ell}_0}=0},
  \eea
   where the component $A^{\ell}_0$ is
defined by the scalar product $A^{\ell}_0=A_\mu {\ell}^\mu=  (A\cdot{\ell})$ of the vector field
$A_\mu$ and the unit time-like vector ${\ell}_\mu$
characterizing the frame of reference  (\ref{3-2a}).

The gauge condition was established by Dirac as the first integral of the Gauss
constraint
 \bea \label{7-2} \int^t dt{\dfrac{\delta W_{\rm QED}}{\delta
A^{\ell}_0}=0},~~~~~~~~ t=(x\cdot \ell).
 \eea
In this case, the S-matrix
elements (\ref{4-2}) are relativistic invariant and independent of the frame
reference provided the condition (\ref{3-2a}) is fulfilled
\cite{Kalinovsky}.

The generalization of the Dirac Hamiltonian approach to QED of 1927 to any frame
  was discussed by  Heisenberg and Pauli's  in 1930 \cite{hp}. It was  their question
   to von Neumann who
   suggested them  to go back to the
  initial Lorentz-invariant formulation  and choose the comoving frame and repeat the gauge invariant
 Dirac scheme in this frame. The comoving frame time axis $\ell_\mu$ (\ref{3-2a}) for a bound state is proportional
 to the total momentum operator \cite{MarkovYukawa,Pervu2}.

The next concept of our scheme is the normal ordering of field operators in this comoving frame.
 It is well known that the  normal ordering of the oscillator
  Hamiltonian $$
{:} \!\!\sum_n \dfrac{p_n^2\! +\! \omega^2_n q^2_n}{2} {:}  =
{:} \sum_n \omega_n \dfrac{a^+_n a^-_n \!+\! a^-_n a^+_n}{2} {:} \! =
\! \sum_n \omega_n \Bigl(\!a^+_n a^-_n \!+\! \textcolor{red}{\dfrac{1}{2}}\!\Bigr)
$$
   leads to the vacuum (Casimir) energy $\sum_n \textcolor{red}{\dfrac{\omega_n}{2}}$ \cite{sh-79}.
The normal ordering of  gluons $A^{a}$ in the SU(3) QCD
including the product of covariant derivatives
$$\nabla^{db}(A)A^b_0\nabla^{dc}(A)A^c_0=
 {:}\nabla^{db}(A)A^b_0\nabla^{dc}(A)A^c_0{:} + M_g^2 A^d_0A^d_0$$ ~
  leads to their effective gluon mass $M_g$ expressed in terms of the condensate $C_{\rm gluon}$:
$$
g^2f^{ba_1d}f^{da_2c}{\langle A^{a_1}_iA^{a_2}_j\rangle} \!
 =\! 3g^2[N_c^2-1] \delta^{bc} \delta_{ij}C_{\rm gluon} = M_g^2 \delta^{bc}\delta_{ij},
$$
where $N_c=3$ and $g^2$ is the QCD coupling constant.
It yields  the Yukawa interaction in QCD
$$
V_{\rm Yu}({\mathbf{k}})=\dfrac{4}{3}g^2\dfrac{1}{\mathbf{k}^2+M_g^2}.
$$
 The ladder diagram sum of Coulomb interaction in QED leads to the Schr\"odinger equation.
In the same way, in QCD, the ladder diagram sum of the Yukawa interaction  between quarks
 leads to the Salpeter equation \cite{a15,Salpeter} and the Schwinger--Dyson one \cite{s-d}:
\be \label{s-d}
\Sigma(k) = m^{0} + {i} \int {dq_0 d^{3}q \over {(2\pi)^{4}}} V_{\rm Yu}( k^{\perp} - q^{\perp} )
\rlap/\ell G_{\Sigma}(q) \rlap/\ell,
\ee
where $ G_{\Sigma}(q) = ( \rlap/q - \Sigma(q))^{-1} $,
$k^{\perp}_{\mu} = k_{\mu}-\ell_{\mu} (k \cdot \ell)$, $\ell^2 = 1$, and $m^0$ is a current mass.
In the reference frame $\ell^{0}=(1,0,0,0)$, $q^{\perp}=(0,\mathbf{q})$ we can put
$$
\Sigma(q) \equiv M({\bf q})=
\sqrt{ M^2 + {\bf q}^2}
 \,\cos  {\upsilon}({\bf q})
$$
and take integration  over $q_0$ in the Schwinger--Dyson equation (\ref{s-d})
$$
 M({\bf q})
=m^0+\frac{1}{2}\int\frac{d^3q}{(2\pi)^3}
V(\textbf{p}-\textbf{q}) \cos {\upsilon}(q) .
$$

 In the limit of small current masses $m_u \simeq m_d\ll M_g$
 and meson ones $M_\pi \to 0$ the S--D equation coincides with the Salpeter one,
 up to the inverse weak-decay coupling constant factor $F_\pi =93$ GeV,
 in agreement with the Goldstone theorem
\bea\label{sd-3u}
{\frac{1}{\sqrt{2}\,F_{\pi}}} \otimes\Bigg|~~~~~~~~~~~~~
{m_d} &=& M_d(p) - \frac{1}{2}\int\frac{d^3q}{(2\pi)^3}
V(\textbf{p}-\textbf{q}) \cos {\upsilon}_d(q), \\
{\frac{M_{\pi}L^{\pi}_{(2)}(\textbf{p})}{2}} &=& \sqrt{p^2+M^2_d(p)}\;~ L^{\pi}_{(1)}
(\textbf{p}) -
\frac{1}{2}\!\int\!\frac{d^3q}{(2\pi)^3}V(\textbf{p}\!-\!\textbf{q}) L^{\pi}_{1}(\textbf{q}), \\
{L^{\pi}_{(1)}(p)} &=&
\frac{M_d(p)}{\sqrt{2}~F_\pi \sqrt{p^2+M^2_d(p)}}={\frac{\cos\upsilon_d(p)}{\sqrt{2}\,F_\pi}}\\
{L^{\pi}_{(2)}(\textbf{p})}
&=&{\frac{2m_d}{\sqrt{2}\,F_\pi\cdot M_{\pi}}}, 
\eea
where $L^{\pi}_{(1)}, L^{\pi}_{(2)}$ are the wave functions
and {$M_d(p)$ is the constituent quark mass}.

 In this case, the normalization condition
$$4N_c\int \dfrac{d^3q}{(2\pi)^3}
 {L^{\pi}_1(q)}{ L^{\pi}_2(q)}=M_\pi$$
 yields the Gell-Mann--Oakes--Renner (GMOR) relation \cite{GellMann:1968rz}
$$
{M^2_{\pi}F^2_{\pi} = {2m_d} <d \bar d>}
$$
between the light quark condensate
$$
<d \bar d>=
\sum_{n=1}^{N_c}\langle q_n(t,{\bf x})\overline{q}_n(t,{\bf x})\rangle =
{4 N_c} \int\frac{d^3q}{(2\pi)^3} \frac{1}{2} \cos\upsilon_u(q),
$$
the current mass $m_d$, the
pion mass ${M_{\pi}}$, and its weak decay coupling constant ${F_{\pi}}$.

In the chiral massless limit ($m^0 \to 0$) the solution of the Schwinger--Dyson equation
 was given in paper by Cherny et al. \cite{5856} in the form of a step-function.
 In the step-function approximation, it was shown in \cite{Kalinovsky} that
 the Schwinger--Dyson equation and the Salpeter one yield meson
  spectrum via the constituent quark masses $M_{\rm const}\simeq 330$ GeV.
Using the GMOR relation and the constituent quark mass value $\sim 330$ MeV
we can define a {\it conformal invariant} as
the ratio of the condensate value to the cubed constituent mass
\be\label{gmor}{\frac{<d\bar d>}{M_d^3}=\frac{M_\pi^2F_{\pi}^2}
{2 m_dM_d^3}\simeq 0.41\pm 0.08
  }.\ee

In the Dirac approach to QCD the color confinement means the complete
 destructive interference of non-Abelian phase factors
 of the topological degeneration of the color states \cite{VP-85},
 so that the ordinary plane wave ${e^{ipx}}$
 is replaced by the sum over parameters of these phase factors
 $$\lim\limits_{L \to \infty} \frac{1}{L}\sum\limits_{n=-L}^{L} \upsilon^{(n)}(x) \,
 \underbrace{{e^{ipx}}}_{{parton}} = 0,$$  if $x\neq 0$.
 It leads to Feynman's quark-hadron duality for $\hat S=1+i \hat T$
 known as parton model. In this case, the left side of the optical theorem
 is the sum over colorless hadron states
\be
 \sum_h <*|\hat T|\underbrace{h><h}_{hadrons}|\hat T|*>=
 \underbrace{{2Im <*|\hat T_{\rm Perturbation~Theory}|*>}}_{partons};
\ee
 whereas the right side of the optical theorem as  the imaginary part
   yields the parton-quark model for all colorless states
marked by stars \cite{VP-85}.
\section{Conformal Version of Minimal Standard Model}

 Now, in the framework of the condensate mechanism of conformal symmetry breaking,
 we consider the conformal version of SM
 (CSM) \cite{cmcsb}.
 For the beginning in CSM one keeps only the scalar field potential part forming
 the four interaction and the largest mass t-quark -- Higgs field interaction
\be \label{hbm1}{{\cal H}^{\rm CSM}_{\rm int}=
\dfrac{\lambda \varphi^4}{4}+\varphi g_t \bar t t},\ee
where {$g_t=1/\sqrt{2}$}.
The t-quark condensate  as a consequence of the normal ordering
$${V_{\rm eff}(\varphi)=\dfrac{\lambda\varphi^4}{4}-\varphi g_t  <t\bar t>}$$
 supersedes the phenomenological
 negative square mass term in the Higgs potential.
 The value of the  t-quark condensate $\bar t t=:\bar t t:-<t\bar t>$
 is estimated using the conformal invariant
  as the ratio  of
 quark condensate and its cubed mass obtained above (\ref{gmor})
$${\frac{<t\bar t>}{M_t^3}=\frac{<d\bar d>}{M_d^3}\simeq0.41\pm0.08
},$$ where {$M_t=173$} GeV  is inputting parameter
yielding the value of the constant part
of the Higgs field $\varphi={{\rm v}}+H$: ${{\rm v}=246}$ GeV.

In  contrast  to the Higgs mechanism,
the condensate mechanism allows us to estimate the coupling constant $\lambda$
\be \label{hbm2}{V'_{\rm eff}({\rm v})=0}~~\Rightarrow~~ {\lambda {\rm v}^3=g_t  <t\bar t>}
 ~~\Rightarrow~~ {\lambda=\frac{<t\bar t>}{4M_t^3}=\frac{0.41\pm0.08}{4}}.\ee
and the
 Higgs boson mass. According to Eqs.(\ref{hbm1}) and (\ref{hbm2}), this mass is in the expected region  
 $${m_{\rm higgs}^{tree}=\sqrt{\frac{3g_t<t\bar t>}{{\rm v}}}=131\pm 15 \mbox{\rm GeV}}.$$

Contributions to this mass from other  electroweak boson condensates
    $<W^+W^->$ and $<ZZ>$ including the Higgs
one $<H\,H>$
are very small \cite{cmcsb}
$$m_{\rm higgs}\simeq m_{\rm higgs}^{\rm tree}\left[1+\frac{\Delta m^2_{\rm higgs}}{m^2_{\rm higgs}}
\right]^{1/2}=m_{\rm higgs}^{\rm tree}[1+0.02]$$
in comparison with the accuracy of the constituent mass definition.

\section{Dirac Conformal  General Relativity and Empty Universe Model}

 Instead of the Hilbert action
$W_{\rm H}=-(1/6)\int d^4x\sqrt{-g}R^{(4)}$ and an interval $ds^2=g_{\mu\nu}dx^\mu dx^\nu$
in  the Riemannian space-time in natural units:
 $
  M_{\rm Pl}\sqrt{3/(8\pi)}=c=\hbar=1
 $
 we consider the Dirac action of the General Relativity \cite{grg,plb10}
 with a scalar dilaton field $D$
$$
 {W_{\rm CGR} = \!\!\!
-\! \! \int\limits_{ }^{ }\!\! \!  d^4x
 \biggl[
  \frac{\sqrt{\! \! -\widetilde{g}}}{6}{R}^{(4)}(\widetilde{g})
 e^{-2D}\!-\!e^{-D}\partial_\mu
 \left(\sqrt{\!-\widetilde{g}}\widetilde{g}^{\mu\nu}\partial_\nu\! e^{-D}\!
\right)\! \biggr]}
 $$
and a
 long distance interval expressed through  the Fock
linear gauge-invariant forms $\widetilde{\omega}^{\rm Fock}_{(\alpha)}$  \cite{Fock:1929vt}
 \be \label{act5}{\widetilde{ds}^2=\widetilde{g}_{\mu\nu}dx^\mu dx^\nu=
 \widetilde{\omega}^{\rm Fock}_{(\alpha)}\otimes\widetilde{\omega}^{\rm Fock}_{(\beta)}\eta^{(\alpha)
 (\beta)}},\ee
where $\eta^{(\alpha)(\beta)}$ is a local tangent Minkowskian space-time metrics.
The frame of reference used for the 4=3+1 foliation of the space-time manifold
 was given by  Dirac \cite{dir2} and Arnowitt--Deser--Misner  \cite{ADM}:
\be\label{act6}
 {\widetilde{\omega}}^{\rm Fock}_{(0)}=e^{-2{D}}N dx^0,~~~~
 {\widetilde{\omega}}^{\rm Fock}_{(b)}={\bf e}_{(b)i}dx^i+{N}_{(b)}dx^0.
\ee
Here $N$ is the lapse function presented as:
$N=N_0(x^0){\cal N}(x^0,x^1,x^2,x^3)$, where $V_0^{-1}\int_{V_0} d^3x {\cal N}^{-1}=1$,
and $V_0=\int d^3x$ is the space volume;
${N}_{(b)}=N^j{\bf e}_{(b)j}$ are the shift vector components;
${\bf e}_{(b)i}$ are the space triad components with unit determinant.

The scalar dilaton field $D$ can be  decomposed over the harmonics
$$ D(x^0,x^1,x^2,x^3)=\langle D\rangle(x^0)+ \overline{D}(x^0,x^1,x^2,x^3),$$
 where $\langle D\rangle(x^0)$ is the   dilaton zero mode and $\overline{D}(x^0,x^1,x^2,x^3)$ is
 the sum of other modes with the constraint $\int_{V_0} d^3x \overline{D}=0$.
 The   dilaton zero mode  plays the role of a reciprocal cosmological scale factor logarithm
\bea \label{tau0}
  \langle D \rangle= - \ln a = \ln(1+z),
\eea
where $z=(1-a)/a$ is the redshift.

 Then, the action is split into the  Newton-type part, graviton part and homogeneous one
 associated with the dynamics of the
 Universe as a whole:
\be \label{act1} W_{\rm CGR}= W_{\rm potential}
 +W_{\rm graviton}^{\rm Fock}+{W_{\rm  Universe}}.\ee

The first part of the action (\ref{act1})

$$ W_{\rm potential}=\int\limits_{}^{}
 d^4x{N}\left[{-{{v^2_{{\overline{D}}}}}}
 -\underbrace{\frac{4}{3}e^{-7D/2}\triangle^{(3)} e^{-D/2}}_{\rm Newtonian~potentials}
 \right],$$
yields the {Newtonian~potentials} in the frame comoving to the local volume element
velocity
${v_{\overline{D}}=0}$.

The second part of the action (\ref{act1})
 $$W^{\rm Fock}_{\rm graviton}=\int\limits_{}^{}
 d^4x\frac{N}{6}\left[{v_{(a)(b)}v_{(a)(b)}}-e^{-4D}R^{(3)}({\bf e})\right]$$
describes the one-component graviton in a squeezed state \cite{plb10}, where the parameter
of squeezing is the dilaton.
{In the Beginning, according to the Bible (Genesis 1:2), the Universe was empty}.
From physical point of view this means that the second and third parts of the action (\ref{act1})
were equal to zero:
\be \label{eum}
W_{\rm graviton}=W_{\rm potential}=0.
\ee

The third part of the action (\ref{act1}) describes the zero mode dynamics of the Universe as a whole:
\bea \label{tau1}
 && W_{\rm Universe}=
 -V_0\int\limits_{\tau_I}^{\tau_0} \underbrace{dx^0 N_0}_{=d\tau}
 \left[\left(\frac{d\, \langle D \rangle}{N_0dx^0}\right)^2
 +{\rho^\tau_{\rm Cas}}\right].
\eea
Here  $d\tau=N_0(x^0)dx^0 $
 is a luminosity time interval connected with the conformal time interval $d\eta$
 and the world one $dt$ by the relations from (\ref{act5}) and (\ref{act6}):
 $$d\tau=a^{-2}d\eta=a^{-3}dt;$$
${\rho^\tau_{\rm Cas}}$
is the {Casimir vacuum energy density} corresponding to the luminosity time interval $\tau$.

In the conformal units the Casimir energy density is able to be represented
as a sum over all physical field energies $\textsf{H}_{\rm Cas}^{(f)}$:
\be\label{vsd}
{\rho^\eta_{\rm Cas}}(a)=
\frac{{\rho^\tau_{\rm Cas}}}{a^{2}}=
\sum_f\frac{\textsf{H}_{\rm Cas}^{(f)}}{V_0}
= \frac{H_0}{d_{\rm Cas}(a)},
\ee
where ${d_{\rm Cas}(a)}$ is the conformal size of the Universe and $H_0$ is the Hubble parameter.

Variations of the action with respect to two independent variables
$\langle D \rangle$ and $N_0$ give the equations of the Empty Universe
\bea \label{tau3}
&&  \frac{\delta W_{\rm Universe}}{\delta\langle D \rangle}=0~~\Rightarrow~~
 2\frac{d }{d\tau}\left[\frac{d \langle D \rangle}{d\tau}\right]
 =\frac{d{\rho^\tau_{\rm Cas}}}{d\langle D \rangle},
\\ \label{tau4}
&& \frac{\delta W_{\rm Universe}}{\delta N_0}=0 ~~\Rightarrow~~
 \left[\frac{d \langle D \rangle}{d\tau}\right]^2={\rho^\tau_{\rm Cas}}.
\eea
The latter equation rewritten in terms of the conformal cosmological factor
$a = \exp (-\langle D \rangle)$
and the conformal density $\rho^\eta_{\rm Cas}(a)$ coincides with the Friedmann
equation
\bea \label{tau5}
\left[\frac{da}{d\eta}\right]^2=\rho^\eta_{\rm Cas}(a).
 \eea
Solution of the Friedmann equation (\ref{tau5}) yields the
conformal  horizon
 \bea \label{tau51}
 d_{\rm horison}(a)=2 r_{\rm horison}(a)=2\int\limits_{0}^{{a}}
 d \overline{a}\;\;[\rho^\eta_{\rm Cas}(\overline{a})]^{-1/2}.
 \eea
The horizon is defined as the distance that a photon covers within
its light cone $d\eta^2-dr^2=0$ for life-time of the Universe. In our case, the horizon
coincides with the visual size of the Universe $d_{\rm Cas}({a})$ in (\ref{vsd}):
\be\label{evs}d_{\rm Cas}(a)=d_{\rm horison}(a).
\ee

The solution of  equations (\ref{tau0}), (\ref{vsd}), (\ref{tau51}), and (\ref{evs})
$$d_{\rm horison}({a}) =\frac{{a^2}}{H_0} ~~~\Rightarrow~~\rho^\tau_{\rm Cas}=H_0^2\equiv\rho_{\rm cr}$$
 yields the Hubble diagram of the description
of Supernova Data in Conformal Cosmology \cite{Behnke:2001nw,Zakharov:2010nf} obtained
 as a
consequence of the Dirac conformal GR in the void space approximation.
In Figure 1 you can see that the {conformal long space interval
$R_{\rm long} = r$}
explains long Supernovae
distances via the dominant Casimir energy (see black line) without the cosmological constant
in the framework of the Empty Universe Model given by (\ref{eum}) and (\ref{evs}).

While the Lambda Cold Dark Matter standard model with
a short space interval $R_{\rm short}  = ra$
requires the $\Lambda$ term dominance to explain
the long Supernovae Distances.

 \begin{center}
\parbox{0.5\textwidth}{
\includegraphics[width=0.5\textwidth,height=0.5\textwidth,clip]{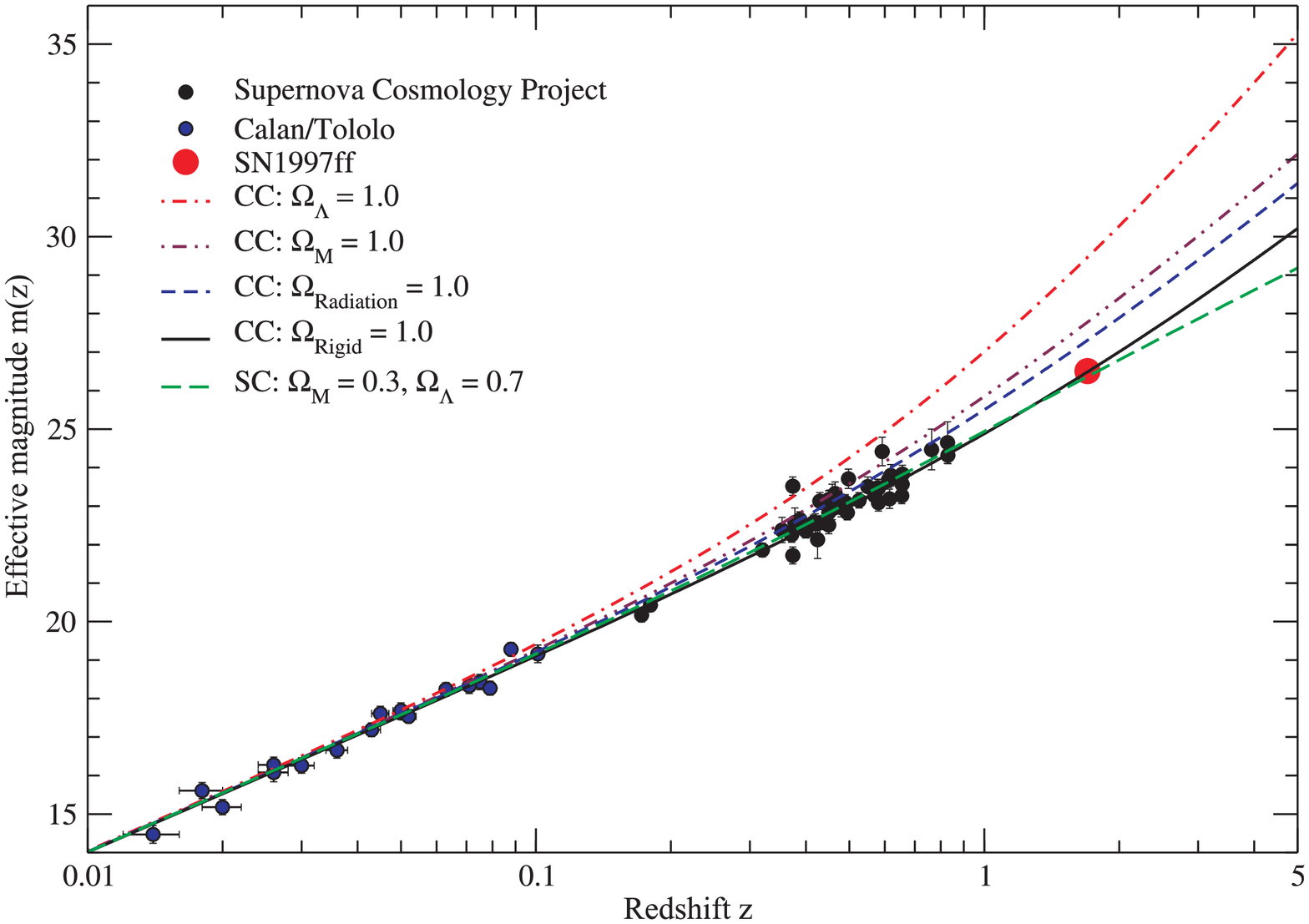}}
\parbox{0.48\textwidth}{
{{The conformal~long~space~interval in the CGR} explains the {\bf long
Supernovae
distances \fbox{$\uparrow$} $R_{\rm SNeIa}$ at $z$ \fbox{$\to$}
via the dominant
{Casimir} energy \cite{Behnke:2001nw,Zakharov:2010nf}
\fbox{$r_{\rm horizon}(z)\!=\!H^{-1}_0(1\!+\!z)^{-2}$}

[see {black line}].\\[.1cm]

\vspace{-0.31cm}

  {The $\Lambda CDM$ model with the short~space~interval}
 requires the $\Lambda$ term dominance to explain the {\bf long Supernovae distances}

$R_{\rm SNeIa}=R_{\Omega_{\Lambda}=0.7,\Omega_{M}=0.3}$

[see {{green line}].
 }
}

\vspace{.31cm}

Fig.1
}}
\end{center}

In paper \cite{grg}, the Empty Universe Model was presented in detail.
It was demonstrated that the Planck least action postulate applied to the
Universe limited by its horizon yields the value of the cosmological
scale factor in the Planck epoch. In other words, the
Planck least action postulate
$$W_{\rm Universe}={\rho_{\rm cr}}V_{\rm hor}^{(4)}(a_{\rm Pl})=
 {M_{\rm  Pl}^2 H_0^2}\frac{1}{H_0^4}\frac{(1+z_{\rm  Pl})^{-8}}{32 }=2\pi
$$
yields the redshift at the Beginning:
 ${a^{-1}_{\rm Pl}=(1+z_{\rm Pl})\simeq 0.85 \cdot 10^{15}}.$
One can see that in the Empty Universe Model the Planck epoch coincides with the electro-weak one.

One can consider, in the tangent space-time, representations
 of the {Weyl group} \cite{Ramond:1984}
 which includes besides of the Poincar\'e group the scale transformations.
This means that the  corresponding  energy
 with respect to the luminosity time interval: ${d\tau} =d\eta/a^2$
takes the form
$${\omega_\tau (a) =a^2\sqrt{{\bf k}^2+a^2M_0^2}}.$$
This energy can be decomposed into different conformal weight parts~~
  $$<\omega_\tau^{(\rm{ n})}(a)>=
 \dfrac{a^{{n}}}{a_{\rm Pl}^{{n}}} H_0$$
responsible for the representations of
the {Weyl group}  in the tangent Minkowskian  space-time.
These representations give scales
$$<\omega_\tau^{(\rm{ n})}(a)>|_{a=1}=\frac{H_0}{a_{\rm Pl}^{{n}}}$$
  for conformal weights {n = 0,\,1,\,2,\,3,\,4} in GeV:

\vspace{.4cm}

\centerline{\begin{tabular}{|c|c|c|c|c|}\hline
 {n}=0&{n}=1&{n}=2&{n}=3&{n}=4 \\ \hline
 $ H_0\!\sim \!10^{-42}$&$R^{-1}_{\rm Celestial~system}\!\sim \!10^{-27}\!$&$
  T_{\rm CMB}\!\sim\! 10^{-12}$&$M_{\rm EW}\!\sim\! 10^{3}$&$M_{\rm Pl}\sqrt{3/(8\pi)}\!\sim\! 10^{18}$\\ \hline
\end{tabular}
}

\vspace{.4cm}

 Thus, conformal weights of the {Weyl group representations}
 with respect to the luminosity  energies give us the present-day $(a=1)$ mass scales  of an order of
 the electroweak scale energy for ${n}=3$,
 a photon energy for ${n}=2$  of an order of  the CMB temperature,
 and
  nonrelativistic energy
$\omega^{(\rm{ 1})}_\tau (a)=a k^2/(2M_0)$ for ${n}=1$ of an order of
the Celestial~system
inverse size.
 The {Weyl group representations} leads to the classification of energy scales
that points out  the common origin of conformal symmetry breaking in both GR and SM.

The intensive creation of primordial gravitons and
Higgs bosons is described assuming that the Casimir vacuum energy is the source
of this process \cite{grg}.

\parbox{0.5\textwidth}{
\includegraphics[width=0.5\textwidth,angle=-0
]{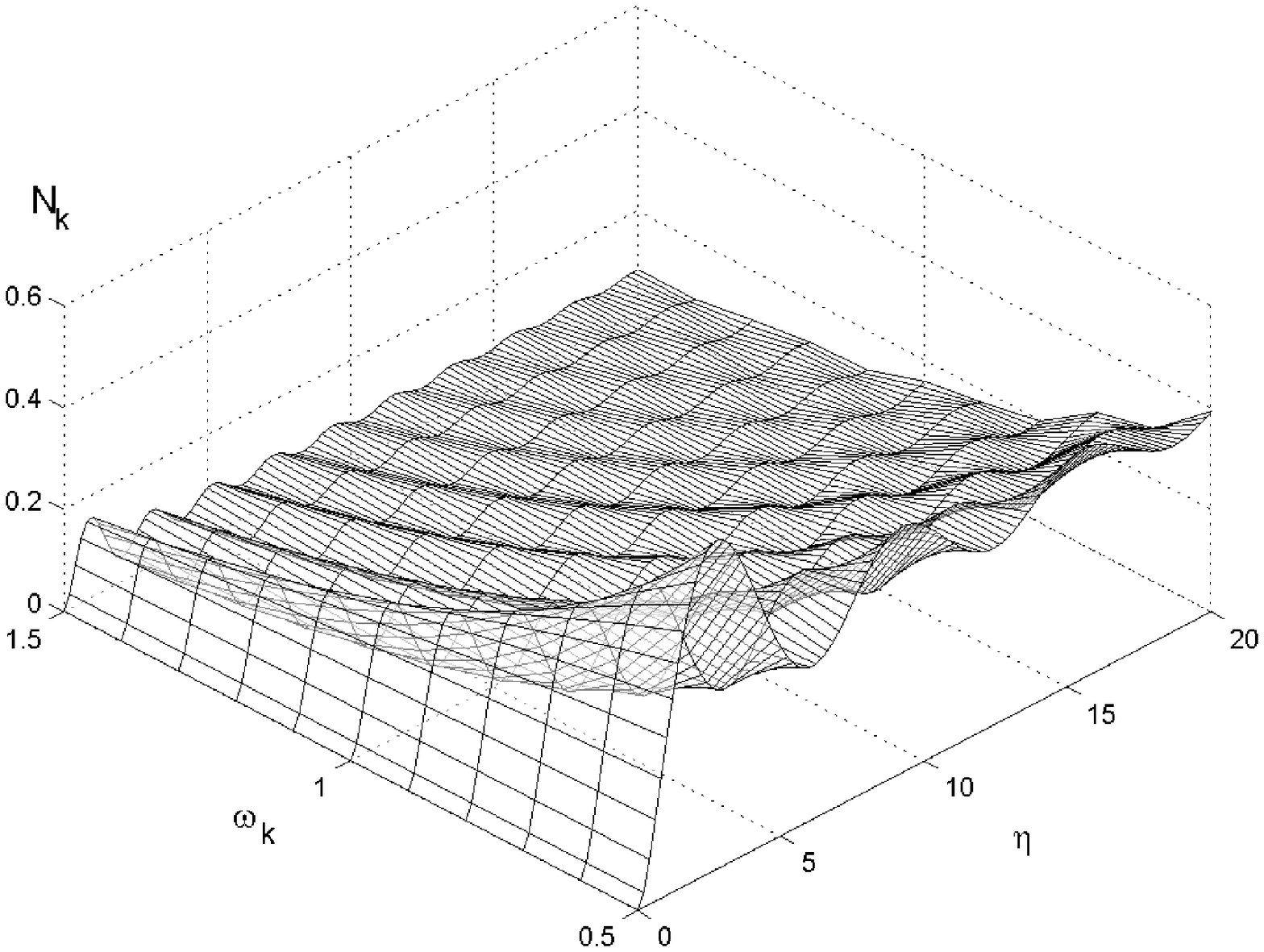}}
\parbox{0.48\textwidth}{

The Casimir energy is the source of creation of $10^{90}$ Higgs particles  \cite{grg}.
This Figure, where
 \fbox{$\nearrow$} is time-axis, 

\fbox{$\uparrow$} is number of bosons $N_{W,Z,h}$,

\fbox{$\searrow$}  is their momentum, shows us creation of $N_h\sim 10^{90}$ Higgs particles  at
{$1+z_{\rm Pl}\sim 10^{15}$} during the first $10^{-12}$sec.
\vspace{.31cm}

 Fig. 2.
}
 The Casimir energy is the source of creation of $10^{90}$ Higgs particles that
decay into $10^{87}$   photons with CMB temperature $\sim$ 3 K in the conformal units.

\section{Conclusion}

At the present talk we tried to demonstrate the mysterious and intrigue fruitfulness of
the conformal symmetry applications: its "footprints" manifest themself on the whole accessible scale --
from quarks to horizons of the Universe. For illustration
we superinduce explicitness in some problems taken from various fields of theoretical physics:
application of the low energy Gell-Mann--Oakes--Renner relation to QCD,
appearance of the Higgs particle mass in Minimal Standard Model of electroweak
interactions without having used the Higgs potential, interpretations of
new observational cosmological data in framework of the Conformal GR.
Everywhere the conformal symmetry breaking is manifested
via the condensate mechanism at the quantum level of description of phenomena.


\newpage
\begin{thebibliography}{99}

\bibitem{Dirac_73}
  P.A.M.~Dirac,
  Proc. Roy. Soc. Lond. A {\bf 333} (1973) 403.

\bibitem{grg}
        V.N.~Pervushin, A.B.~Arbuzov, B.M.~Barbashov, R.G.~Nazmitdinov,
        A.~Borowiec, K.N.~Pichugin, and A.F.~Zakharov,
        Gen. Relativ. Gravit.
        (2012) DOI 10.1007/s10714-012-1423-7.

\bibitem{a15} E.E.~Salpeter and H.A.~Bethe, Phys. Rev. {\bf 84} (1951) 1232.

\bibitem{Salpeter}
E.E.~Salpeter, Phys. Rev. D \textbf{87} (1952) 328.

\bibitem{MarkovYukawa}
M.A.~Markov, J. Phys. (USSR) \textbf{3} (1940) 453;
H.~Yukawa, Phys. Rev. \textbf{77} (1950) 219.

\bibitem{Pervu2}
V.N.~Pervushin, Nucl. Phys. B (Proc. Supp.) \textbf{15} (1990) 197.

\bibitem{Dirac}
P.A.M.~Dirac, Proc. Roy. Soc. A \textbf{114} (1927) 243;
Can. J. Phys. \textbf{33} (1955) 650.

\bibitem{logunov}
N.N.~Bogoliubov, A.A.~Logunov, A.I.~Oksak, and I.T.~Todorov,
{\it General Principles of Quantum Field Theory}, Springer (1989).

\bibitem{Kalinovsky}
Yu.L.~Kalinovsky, L.~Kaschluhn, and V.N.~Pervushin, Phys. Lett. B \textbf{231} (1989) 288;
Fortsch. Phys. \textbf{38} (1990) 353.

\bibitem{hp}
W.~Heisenberg and W.~Pauli, Z. Phys. \textbf{56} (1929) 1; Z. Phys. \textbf{59} (1930) 166.

\bibitem{sh-79}
  A.A.~Actor,
  Fortsch. Phys. {\bf 43} (1995) 141; \\
  M.~Bordag, G.L.~Klimchitskaya, U.~Mohideen, and V.M.~Mostepanenko,
  {\it Advances in the Casimir Effect},
  Oxford University Press Inc., New York (2009).

\bibitem{s-d} F.~Dyson, Phys. Rev., \textbf{75} (1949) 1736;
J.~Schwinger, PNAS \textbf{37} (1951) 452.


\bibitem{GellMann:1968rz}
       M.~Gell-Mann, R.J.~Oakes, and B.~Renner,
       Phys.\ Rev.\ {\bf 175} (1968) 2195.

\bibitem{5856} A.Yu.~Cherny, A.E.~Dorokhov, Nguyen Suan Han,
 V.N.~Pervushin, V.I.~Shilin,
Bound States in Gauge Theories as the Poincar\'e Group Representations, arXiv:1112.5856 [hep-th].

\bibitem{VP-85} V. Pervushin, {Riv. del Nuovo Cimento} {\bf 8} (1985) N 10,  1--48.

\bibitem{cmcsb} V.N.~Pervushin, A.B.~Arbuzov, R.G.~Nazmitdinov, A.E.~Pavlov, A.F.~Zakharov,
Condensate Mechanism of Conformal Symmetry Breaking and the Higgs Boson, arXiv:1209.4460 [hep-ph].

\bibitem{plb10}
A.B.~Arbuzov, B.M.~Barbashov, R.G.~Nazmitdinov, V.N.~Pervushin, A.~Borowiec, K.N.~Pichugin, A.F.~Zakharov,
 Phys. Lett. B {\bf 691} (2010)  230;
[{arXiv:1007.0293} [gr-qc]].

\bibitem{Fock:1929vt}
  V.~Fock,
  Z. Phys. {\bf 57} (1929) 261.

\bibitem{dir2}
   P.A.M.~Dirac,
   Proc. Roy. Soc. Lond. A {\bf 246} (1958) 333.

\bibitem{ADM}
   R.~Arnowitt, S.~Deser, and C.W.~Misner,
   {\it The dynamics of general relativity,}
   in L.~Witten,
   {\it  Gravitation: An Introduction to Current Research,}
   Wiley, New York (1962) p.227.

\bibitem{Behnke:2001nw}
   D.~Behnke, D.B.~Blaschke, V.N.~Pervushin and D.~Proskurin,
   Phys. Lett. B {\bf 530} (2002) 20;
   [arXiv:gr-qc/0102039].

\bibitem{Zakharov:2010nf}
   A.F.~Zakharov and V.N.~Pervushin,
   Int. J. Mod. Phys. D {\bf 19} (2010) 1875;
   [arXiv:1006.4745 [gr-qc]].

\bibitem{Ramond:1984}
P.~Ramond, {\it Field Theory. A Modern Primer,} The Benjamin / Cummings Publishing Company, Inc. (1981).

\end{thebibliography}
\end{document}